\documentclass[a4paper,11pt]{article}
\usepackage{subcaption}
\usepackage{epsfig, graphicx, subfloat, float}
\usepackage{latexsym}
\usepackage{cite}
\usepackage{amssymb}
\usepackage{latexsym}
\usepackage{mathrsfs}
\usepackage{amsmath}
\usepackage{color}
\usepackage[usenames]{xcolor}
\graphicspath{{Image/}}
\author{H. Mohseni Sadjadi \footnote{mohsenisad@ut.ac.ir}
\\ {\small Department of Physics, University of Tehran,}
\\ {\small P.O.B. 14395-547, Tehran 14399-55961, Iran}}
\title {Onset of dark energy from cosmological scalarization and gravitational wave speed }
\begin{document}
\maketitle
\begin{abstract}
 This study investigates the emergence of dark energy during the matter-dominated era through spontaneous cosmological scalarization
 in the scalar-Ricci-Gauss-Bonnet model.Our model aligns with the conjecture that the speed of gravitational waves at low
 redshifts is nearly equal to the speed of light, implying that the Gauss-Bonnet invariant does not directly contribute to late-time
 acceleration and suggesting that scalarization begins in the radiation-dominated era. The Gauss-Bonnet coupling stabilizes the quintessence
 field at an initial fixed point with negligible dark energy density during the radiation era.
  As the Universe expands, the Ricci coupling induces a tachyonic transition in the quintessence field, driving the evolution of dark energy.

\end{abstract}
\section{Introduction}

Scalar-tensor models extend General Relativity (GR) by incorporating scalar fields alongside the geometric tensor fields of GR.
 These models are frequently invoked to address cosmological and astrophysical phenomena that GR alone does not fully explain,
 such as the accelerated expansion of the late-time Universe \cite{Acc1,Acc2,Acc3,Acc4,Acc5,Acc6,Acc7}.
 In scalar-tensor models, a scalar field—often associated with quintessence—may couple to curvature invariants like the
 Ricci scalar \cite{RS1,RS2,RS3} or the Gauss-Bonnet (GB) term \cite{GB1,GB2,GB3,GB4,GB5,GB6,GB7,GB8,GB9,GB10,GB11,GB12,GB13,GB14,GB15,GB16,GB17,
 GB18},
 which has roots in the low-energy limit of string theory \cite{Gross, GB3}.

A particularly intriguing feature of these models is scalarization, a phenomenon where the scalar field’s effective potential,
 shaped by its couplings, allows non-trivial scalar field configurations that deviate from GR.
 This is typically triggered by an instability when the effective mass of the scalar field becomes tachyonic,
 leading to significant deviations in the field's behavior. Scalarization has been studied extensively
 in the contexts of neutron stars \cite{SC1} and black holes \cite{SC2}.

Dark energy, which dilutes more slowly than matter and radiation due to redshift, would be expected to eventually
 dominate the Universe entirely. However, current observations indicate that dark energy and matter are of the same
  order of magnitude today, posing the "coincidence problem" \cite{coin1,coin2,coin3,coin4,coin5}.
 One possible resolution is that dynamical dark energy was initially negligible but became significant during
 the matter-dominated era. Observations, particularly those related to structure formation, also suggest that
 the amount of dark energy in the early Universe was relatively small or even negligible (e.g., \cite{Doran,Bartelmann,Hill,Gomez}).
 This motivates the need for a theoretical framework that explains why dynamical dark energy was negligible during the early era.

Additionally, understanding the interplay between scalar-tensor models and constraints on the speed of gravitational waves offers
new insights into the nature of dark energy and the underlying theory of gravity. In scalar-tensor models,
the gravitational wave speed (GWS) typically deviates from that of light, and this deviation is constrained
by observational bounds. Such constraints have ruled out many models as viable explanations for
the late-time acceleration of the Universe. A pivotal event was the detection of gravitational waves from the binary neutron
star merger GW170817, nearly simultaneously with its electromagnetic counterpart, observed by LIGO, Virgo,
and several telescopes \cite{ob1,ob2,ob3}. This event placed stringent limits on the deviation of the GWS from the speed of light,
showing that at low redshift ($z<0.009$), the deviation is smaller than one part in $10^{15}$ \cite{ob4}.
This result has profound implications, excluding significant portions of parameter space in many modified gravity theories \cite{doub}.

In this paper, within a scalar-tensor model by exploring the connection between three aforementioned seemingly unrelated topics,
i.e. scalarization,
negligible early dark energy, and the GWS constraints, we study  the emergence of dark energy in matter dominated era.
We consider a scenario where a quintessence
field is coupled to both the Ricci scalar and the GB term. The requirement that the GWS is nearly equal to the speed of light
implies that the GB term cannot directly influence the Universe's late-time evolution but instead establishes conditions
for a stable fixed point with negligible dark energy density during the radiation-dominated era. As the Universe expands,
the coupling to the Ricci scalar destabilizes this fixed point, triggering a spontaneous cosmological scalarization mechanism
that gives rise to dark energy. For a quadratic power-law coupling of the quintessence field to the GB term, the GWS's proximity
to the speed of light at low redshifts suggests that scalarization begins in the radiation-dominated era but due to the friction term
becomes marginal during the matter-dominated era.

The structure of this paper is as follows: In Section 2, we provide the necessary preliminaries. In Section 3, we examine
the impact of the GWS constraint on the late-time acceleration within our nonminimal coupling model. In section 4, we explain and formulate
onset of dark energy based on spontaneous scalarization by considering the GWS constraints, offering numerical illustrations using a
massive scalar field model with non-minimal quadratic couplings. Before concluding we add a short note about parameterized
post Newtonian formalism of the model. Finally, in Section 5, we summarize our main results.

Throughout this paper, we use natural units where $\hbar = c = 1$.

 \section{preliminaries}
We begin with the action
\begin{equation}\label{1}
S=\int d^4x \sqrt{-g}\left(\frac{1}{2}uR-\frac{1}{2}g^{\mu \nu}\partial_\mu \phi \partial_\nu \phi-V-\frac{1}{2}f\mathcal{G}\right)+S_m(\Psi_m),
\end{equation}
$u=u(\phi)$, $f=f(\phi)$, and $V=V(\phi)$ are functions of the quintessence scalar field $\phi$. $R$
 is the Ricci scalar and the Gauss-Bonnet (GB) invariant $\mathcal{G}$, is defined as
\begin{equation}\label{0}
\mathcal{G}=R_{\mu \nu \rho \sigma}R^{\mu \nu \rho \sigma}-4R_{\mu \nu}R^{\mu \nu}+R^2
\end{equation}
$S_m$ is the matter action including other ingredients  $\Psi_m$ with barotropic equation of state (EoS),
 such as pressureless dark and baryonic matter, and radiation.
 We consider the spatially flat Friedmann-Lema\^{i}tre-Robertson-Walker (FLRW) space-time with the metric
\begin{equation}\label{2}
ds^2=-dt^2+a^2(t)(dx^2+dy^2+dz^2),
\end{equation}
where $a(t)$ is the scale factor. We scale $a=1$ for our present time. In terms of the Hubble parameter $H=\frac{\dot{a}}{a}$,
 the Ricci scalar and the GB invariant are
\begin{eqnarray}\label{3}
R=6(\dot{H}+2H^2)\nonumber \\
\mathcal{G}=24H^2(\dot{H}+H^2).
\end{eqnarray}
Modified Friedmann equations are obtained as
\begin{eqnarray}\label{4}
&&3(u-4H\dot{f})H^2=\frac{1}{2}\dot{\phi}^2+V-3H
\dot{u}+\rho_m+\rho_r \nonumber \\
&&2(u-4H\dot{f})\dot{H}=-\dot{\phi}^2-\ddot{u}+H\dot{u}+4H^2(\ddot{f}-H\dot{f})-\rho_m-\frac{4}{3}\rho_r
\end{eqnarray}
$\rho_m$ and $\rho_r$ are cold matter (including baryonic and dark matter) and radiation densities, respectively,
satisfying the continuity equations
\begin{eqnarray}\label{5}
\dot{\rho_m}+3H\rho_m=0\nonumber \\
\dot{\rho_r}+4H\rho_r=0
\end{eqnarray}
whose solutions are $\rho_m(a)=\rho_m (a=1) a^{-3}$ and $\rho_r(a)=\rho_r(a=1)a^{-4}$.

We may rewrite the modified Friedmann equation as the standard one by including a dark energy component
\begin{eqnarray}\label{7}
H^2=\frac{1}{3M_P^2}(\rho_m+\rho_r+\rho_D)\nonumber\\
\dot{H}=-\frac{1}{2M_P^2}(\rho_m+\frac{4}{3}\rho_r+P_D+\rho_D)
\end{eqnarray}
where $M_P$ is the reduced Planck mass, and
\begin{equation}\label{8}
\rho_D=\frac{1}{2}\dot{\phi}^2+V+3H^2(M_P^2-u)+3H(4H^2\dot{f}-\dot{u})
\end{equation}
is a hypothetical dark energy density whose pressure is
\begin{equation}\label{9}
P_D=\frac{1}{2}\dot{\phi}^2-V-3H^2(M_P^2-u)-8H^3\dot{f}+2H\dot{u}+2(u-4H\dot{f}-M_P^2)\dot{H}+\ddot{u}-4H^2\ddot{f}
\end{equation}
The dark energy satisfies the continuity equation:
\begin{equation}\label{10}
\dot{\rho}_D+3H(P_D+\rho_D)=0
\end{equation}
 If we respect the nonnegativity energy density condition for dark energy, this interpretation ((\ref{8}) and (\ref{9}))
 is valid as long as  $\rho_D\geq 0$.

\section{GB invariant, GWS and the late time acceleration}

The model (\ref{1}) belongs to the Horndeski's models which describe the most general scalar-tensor theory
 with second-order field equations \cite{Hor}
\begin{eqnarray}\label{11}
&&S=\int d^4x \sqrt{-g} \Big[G_2-G_3\Box \phi(x)+G_4R+G_{4,X}\left((\Box \phi)^2-(\nabla_\mu \nabla_\nu \phi)(\nabla^\mu \nabla^\nu \phi)\right)\nonumber \\
&&-\frac{1}{6}G_{5,X}\left((\Box \phi)^3-3(\Box \phi)(\nabla_\mu \nabla_\nu \phi)(\nabla^\mu \nabla^\nu \phi)+
2(\nabla^\mu \nabla_\alpha \phi)(\nabla^\alpha \nabla_\beta \phi)(\nabla^\beta \nabla_\mu \phi)\right)\Big]\nonumber \\
&&+G_5G_{\mu \nu}\nabla^\mu \nabla^\nu \phi+S_m
\end{eqnarray}
where $X=-\frac{1}{2}g^{\mu \nu}\nabla_\mu \phi \nabla_\nu \phi$, and $G_{\mu \nu}$ is the Einstein tensor.
$G_i$ are assumed to be functions of $\phi$ and $X$. By some computation we find out that the model (\ref{1}) may be resulted
in from (\ref{11}) by choosing
\begin{eqnarray}\label{12}
&&G_2=X-V-4f_{,\phi\phi\phi\phi}X^2(3-lnX)\nonumber \\
&&G_3=-2f_{,\phi\phi\phi}X(7-3\ln X)\nonumber\\
&&G_4=\frac{u}{2}-2f_{,\phi\phi}X(2-\ln X)\nonumber \\
&&G_5=2f_{,\phi} \ln X.
\end{eqnarray}
A comma subscript means a partial derivative. In Horndeski's models, the speed of gravitational waves is generally different from
the speed of light and is \cite{stab5,stab6}
\begin{equation}\label{13}
c_{T}^2=\frac{G_4-XG_{5,\phi}-XG_{5,X}\ddot{\phi}}{G_4-2XG_{4,X}-XG_{5,X}\dot{\phi}H+XG_{5,\phi}}.
\end{equation}
For our model, this reduces to (to be more clear, we insert the light speed $c$ in this formula)
\begin{equation}\label{14}
\frac{c_{T}^2}{c^2}=\frac{4f_{,\phi\phi}\dot{\phi}^2+4f_{,\phi}\ddot{\phi}-u}{4Hf_{,\phi}\dot{\phi}-u}=\frac{4\ddot{f}-u}{4H\dot{f}-u}.
\end{equation}

So the assumption $\frac{c_{T}^2}{c^2}\simeq 1$ restricts the coupling functions and the parameters of the model. In \cite{ob4}, based on
the detection of gravitational wave (GW170817 event) due to the merger of two neutron stars, together with its electromagnetic counterpart,
 it is claimed that at the late time $z<0.009$ we have:
\begin{equation}\label{15}
-3\times 10^{-15}\leq \frac{c_{T}}{c}-1 \leq 7\times 10^{-16}.
 \end{equation}
Respecting this bound requires that $\ddot{f}\ll u$ and $4H\dot{f}\ll u$. In this situation, from modified Friedmann equations (\ref{4})
we conclude that the GB term has no direct influence on the late time acceleration.
However, the GB may play another role in earlier epochs, as we will explain in the following subsection.

\section{Scalarization and onset of dark energy}

The equation of motion of the homogenous scalar field (quintessence) is
\begin{equation}\label{6}
\ddot{\phi}+3H\dot{\phi}+V^{eff.}_{,\phi}=0.
\end{equation}
where $V^{eff.}_{,\phi}=V_{,\phi}+(H^2+\dot{H})(12H^2f_{,\phi}-3u_{,\phi})-3H^2u_{,\phi}$.
So, the quintessence evolution depends on the coupling to the Ricci scalar and the GB terms and, through them
(and the friction term) to other ingredient's densities. This provides the scalarization scenario.
In our model construction we assume that $V$, $u$, and $f$ are even functions of the quintessence and consequently (\ref{1})
has a $Z_2$ symmetry (under $\phi \to -\phi$). As a result we have
$V_{,\phi}(0)=u_{,\phi}(0)=f_{,\phi}(0)=0$ leading to $V^{eff.}_{,\phi}(0)=0$.
Therefore $\phi=0$ is a trivial solution to the equation (\ref{6}) respecting the
$Z_2$ symmetry of the action. We assume also $u(0)=M_P^2$, and $V(0)=0$. Therefore dark energy density (\ref{8}) vanishes for this
trivial solution and the model reduces
to an ordinary GR model satisfying the following ordinary
Friedmann equations, without any modification:
\begin{eqnarray}\label{16}
3M_P^2H^2=\rho_m+\rho_r\nonumber \\
2M_P^2\dot{H}=-\rho_m-\frac{4}{3}\rho_r.
\end{eqnarray}
So we may have a stable solution with negligible dark energy density. We aim to attribute this phase
to era before the onset of dark energy which occurs later due to the scalarization in which
the system gains a nontrivial quintessence solution and deviates from GR. By considering small fluctuation around $\phi=0$,
the quintessence effective mass squared is derived from (\ref{6}) as
\begin{eqnarray}\label{17}
&&m_{eff.}^2(\phi=0)=V^{eff.}_{,\phi \phi}(0)=-\frac{2}{3M_P^4}(\rho_m+\rho_r)(\rho_m+2\rho_r)f_{,\phi\phi}(0)\nonumber \\
&&-\frac{1}{2M_P^2}\rho_m u_{,\phi\phi}(0)
+V_{,\phi\phi}(0).
\end{eqnarray}

By taking $f_{,\phi\phi}(0)<0$, we see that $m_{eff.}^2(\phi=0)>0$, and $\phi=0$ is a stable point for large radiation or
matter densities (early eras). So the GB coupling may provide conditions to have a negligible dark energy density phase in early eras.
For a cosmological scalarization to occur we require that the quintessence becomes tachyonic
$m_{eff.}^2(\phi=0)<0$. This can be accomplished though the signature changing of $m_{eff.}^2(\phi=0)$ in (\ref{17}),
during the Universe expansion, by densities redshifts. For this, we require that at least one of the remaining two terms
of the right hand side of (\ref{17}) be negative ($V_{,\phi\phi}(0)<0$ or $u_{,\phi\phi}(0)>0$) for $a>a_c$, where the critical
scale factor $a_c$ is defined as $m_{eff.}(a_c)=0$, and $\frac{dm_{eff.}^2}{da}(a_c)<0$. Note that if we did not consider
the Ricci coupling, $V_{,\phi\phi}(0)$
would be negative, which, with the assumption $V(0)=0$, would lead to a negative potential for $a>a_c$
which could not lead to an accelerated expansion \cite{RS1}. After the square of the effective mass becomes negative, $\phi=0$
becomes an unstable point, and the quintessence evolution begins and the dark energy emerges. The system gains a nontrivial solution
deviating from GR, satisfying (\ref{7}). A non-zero potential emerges which is required for the late time acceleration \cite{RS1}.

An intriguing aspect of (\ref{17}) is that, for a quadratic coupling $f=-\frac{1}{2}\alpha^2 \phi^2$,
the value of $\alpha$ required to respect the assumption (\ref{15}) is obtained when $a_c$ is
in the radiation dominated era. However, as we will see due to the friction term, the dark energy
actually rises in the matter-dominated era.

Note that residing off the scalar field at $\phi=0$ prevents possible instabilities
reported for the GB model in the inflationary era \cite{early}.

To illustrate our results more specifically, let us proceed with a particular model. Consider a massive scalar field with
simple quadratic non-minimal couplings (as our primary focus is on modifying the quintessence effective mass):
\begin{equation}\label{18}
u=M_P^2+\frac{1}{2}\epsilon^2 \phi^2, \,\, f=-\frac{1}{2}\alpha^2\phi^2,\,\,V=\frac{1}{2}\mu^2\phi^2,
\end{equation}
where $\alpha$, $\epsilon$ and $\mu$ are real numbers. For $\alpha=\epsilon=0$, we recover the minimal model with a massive scalar field.
For our numerical analysis we use dimensionless parameters:
$\hat{\phi}=\frac{\phi}{M_P},\,\, \hat{H}=\frac{H}{H^*},\,\, \hat{t}=H^*t,\,\, \hat{\mu}= \frac{\mu}{H*},
\,\,\, \hat{\alpha}=H^*\alpha,\,\,\, \hat{\rho}=\frac{\rho}{M_P^2H^{*2}}$, where $H^*$ is an energy scale.
In terms of these parameters, the evolution equations may be rewritten as
\begin{eqnarray}\label{21}
&&\hat{\phi}''+3\hat{H}\hat{\phi}'+\hat{\mu}^2\hat{\phi}-(\hat{H}^2+\hat{H}')(3\epsilon^2\hat{\phi}
+12\hat{\alpha}^2\hat{H}^2 \hat{\phi})-3\epsilon^2\hat{H}^2\hat{\phi}=0\nonumber \\
&&3(1+\frac{1}{2}\epsilon^2\hat{\phi}^2)\hat{H}^2=\frac{1}{2} \hat{\phi}'^2+\frac{1}{2}\hat{\mu}^2\hat{\phi}^2-12\hat{\alpha}^2\hat{H}^3\hat{\phi}\hat{\phi}'
-3\epsilon^2\hat{H}\hat{\phi}\hat{\phi}'+\hat{\rho}_m+\hat{\rho}_r\nonumber \\
&&\hat{\rho}_m'+3\hat{H}\hat{\rho}_m=0\nonumber \\
&&\hat{\rho}_r'+4\hat{H}\hat{\rho}_r=0,
\end{eqnarray}
where the prime denotes derivative with respect to the dimensionless time $\hat{t}=H^*t$.

In our model we require that the system has a trivial stable quintessence solution $\hat{\phi}=0$ for $a<a_c$, but that the effective mass
squared becomes negative for $a\geq a_c$, making $\hat{\phi}=0$ an unstable point. The GS coupling parameter may be expressed in terms of $a_c$
as:
\begin{equation}\label{222}
\hat{\alpha}^2=\frac{\frac{1}{2}\epsilon^2\hat{\rho}_{m0}a_c^{-3}-\hat{\mu}^2}
{\frac{2}{3}(\hat{\rho}_{m0}a_c^{-3}+\hat{\rho}_{r0}a_c^{-4})(\hat{\rho}_{m0}a_c^{-3}+2\hat{\rho}_{r0}a_c^{-4})}.
\end{equation}
Smaller values of $a_c$ yield smaller values for $\hat{\alpha}^2$, which favor the proximity of gravitational wave speed (GWS)
to the speed of light, $\lim_{\hat {\alpha} \to 0}\left(\frac{c_T}{c}\right)^2 = 1$, as can be seen from:
\begin{equation}\label{24}
\left(\frac{c_T}{c}\right)^2=
\frac{8\hat{\alpha}^2\hat{\phi} \hat{\phi}''+8\hat{\alpha}^2\hat{\phi}'^2+2+
\epsilon^2\hat{\phi^2}}{8\hat{\alpha}^2\hat{H}\hat{\phi}\hat{\phi}'+2+\epsilon^2 \hat{\phi}^2}.
\end{equation}
Thus, we take $a_c$ in the radiation dominated era. However, due to the friction term, scalarization in (\ref{6}) does not occur immediately.
 To show the evolution of the system, we use equations (\ref{21}), and choose the boundary conditions
 at $a=1/37500$ (in radiation dominated era), as
\begin{equation}\label{22}
\hat{\phi}=10^{-60},\hat{\phi}'=10^{-60},\hat{\rho}_m=25\times 10^{10},\hat{\rho}_r=25\times 10^{11}.
\end{equation}

\begin{figure}[H]
\centering
\includegraphics[height=6cm]{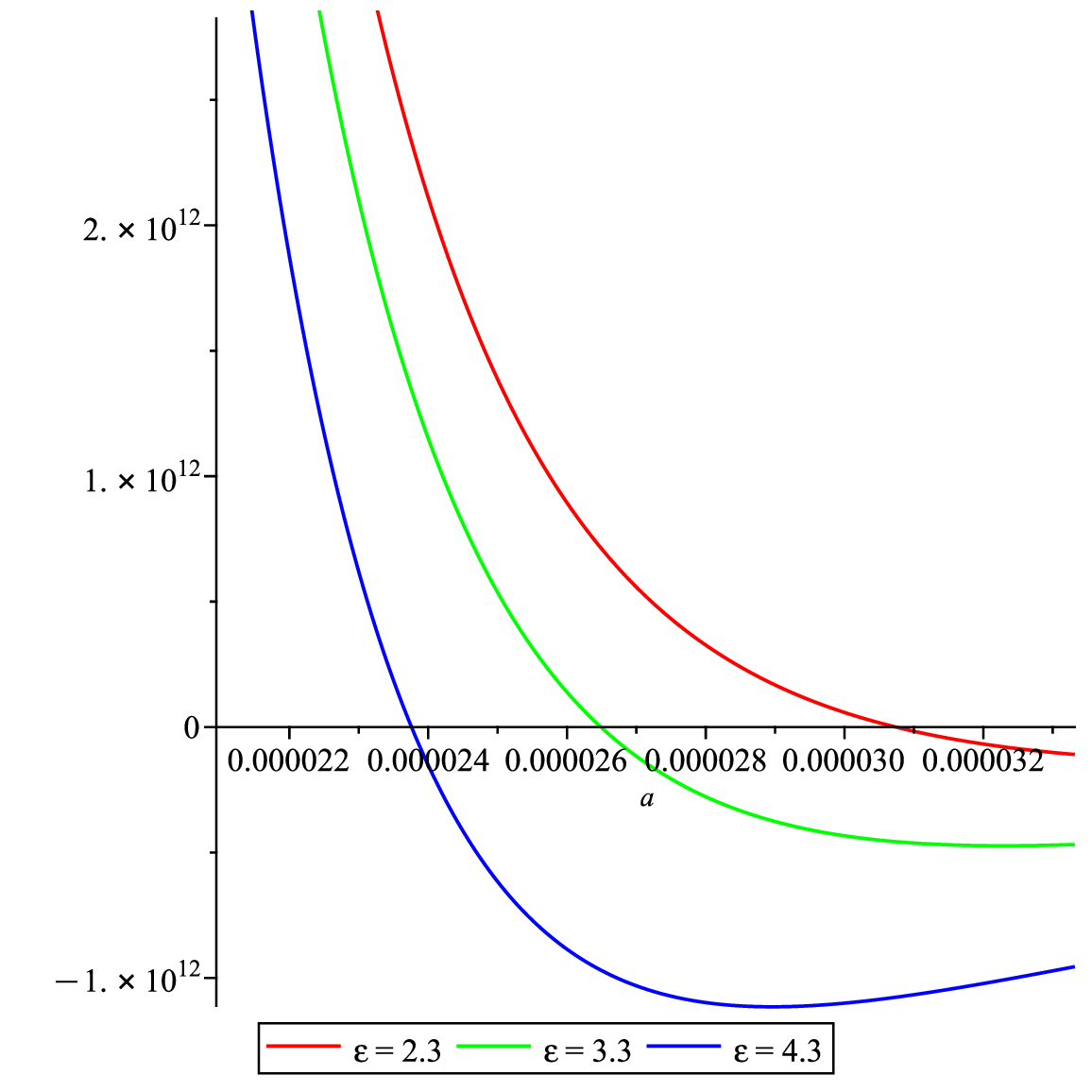}
\caption{The quintessence effective mass squared in terms of the scale factor.}
\label{fig0}
\end{figure}

\begin{figure}[H]
\centering
\includegraphics[height=6cm]{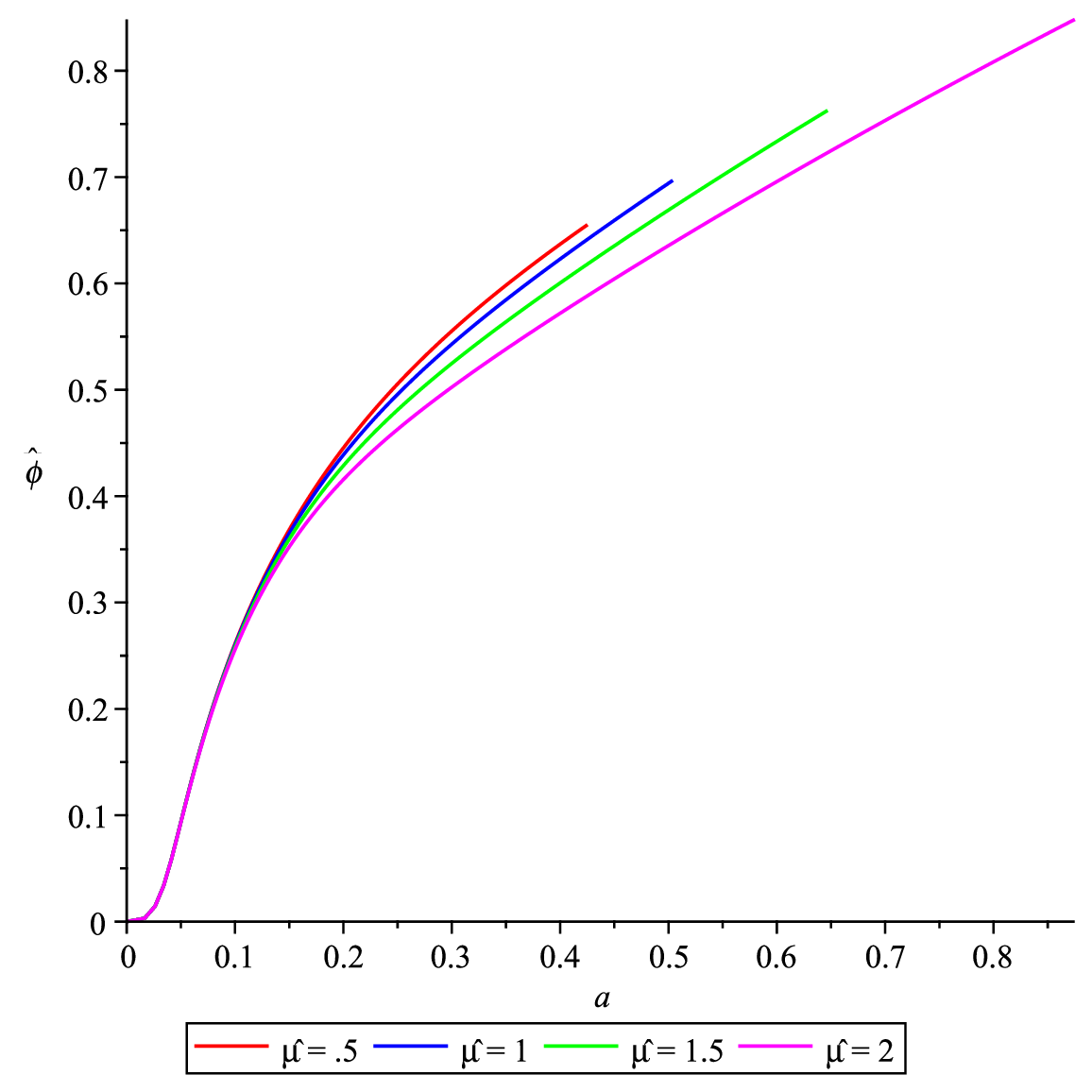}
\caption{The quintessence evolution in terms of the scale factor.}
\label{fig1}
\end{figure}

By using (\ref{21}) and choosing $\hat{\alpha}=3.76 \times 10^{-7},\,\,\, \hat{\mu}=0.5$,
the dimensionless effective mass squared, $\frac{m_{eff.}^2}{{H^*}^2}$, is depicted in fig.(\ref{fig0}) in terms of the scale factor for some $\epsilon$.
This figure shows that for scale factors bigger than a critical value $a_c$, the quintessence becomes tachyonic.
For larger $\epsilon$ this occurs at smaller $a_c$.
When the quintessence effective mass squared becomes negative, the instability triggers scalarization, leading to a nontrivial solution for
quintessence. Based on (\ref{21}), (\ref{22}), and by choosing $\hat{\alpha}=3.76 \times 10^{-7},\,\, \epsilon=3.3$ ,
we illustrate the behavior of the quintessence in terms of the scale factor in fig.(\ref{fig1}) for some values of $\mu$.
The quintessence evolves from $\phi=0$ after $a_c=1/37500$; the evolution depends on densities and
the friction term. For different values of $\hat{\mu}$, the quintessence behaves differently at late times. For smaller values of $\hat{\mu}$,
the slope of the scalar field is steeper. Due to the friction term, the quintessence remains
at $\hat{\phi}\approx 0$ until the matter dominated era. For example for $\hat{\mu}=0.5$, we have we have $\hat{\phi}\simeq0.001$.

\begin{figure}[H]
\centering
\includegraphics[height=6cm]{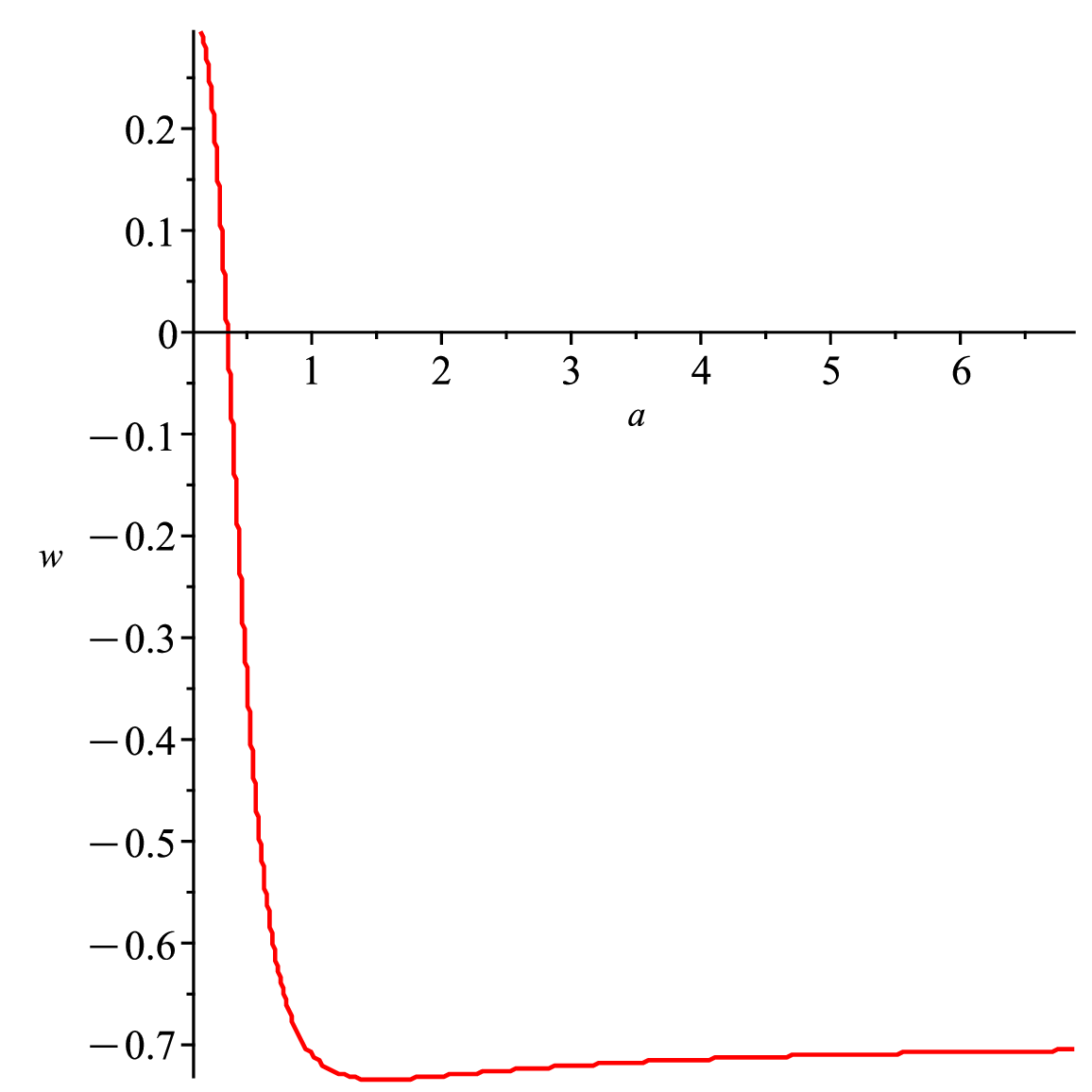}
\caption{The Universe equation of state parameter in terms of the scale factor.}
\label{fig2}
\end{figure}

With scalarization, the Friedmann equations are modified and deviate from GR.
The equation of state parameter of the Universe,
 $w=-1-\frac{2}{3}\frac{\dot{H}}{H^2}$, is plotted in fig.(\ref{fig2}) for the conditions in (\ref{22}) and parameters chosen as
 \begin{equation}\label{23}
\hat{\alpha}=3.76 \times 10^{-7},\,\,\, \hat{\mu}=0.5,\,\, \epsilon=3.3,
\end{equation}
corresponding to $a_c=1/37500$. Hereafter, we use parameters from (\ref{23}) and boundary conditions from (\ref{22}) set at $a = a_c = 1/37500$.
(Setting $H(a=1)=H_0$ as the present Hubble parameter, $H^*$ is obtained as $H^*\simeq 14.3 H_0$. Therefore $\hat{\mu}=0.5$ in (\ref{23}) yields
 the mass of the quintessence as the same order of magnitude as $H_0$: $\mu\simeq 7.15 H_0$).

\begin{figure}[H]
\centering
\includegraphics[height=6cm]{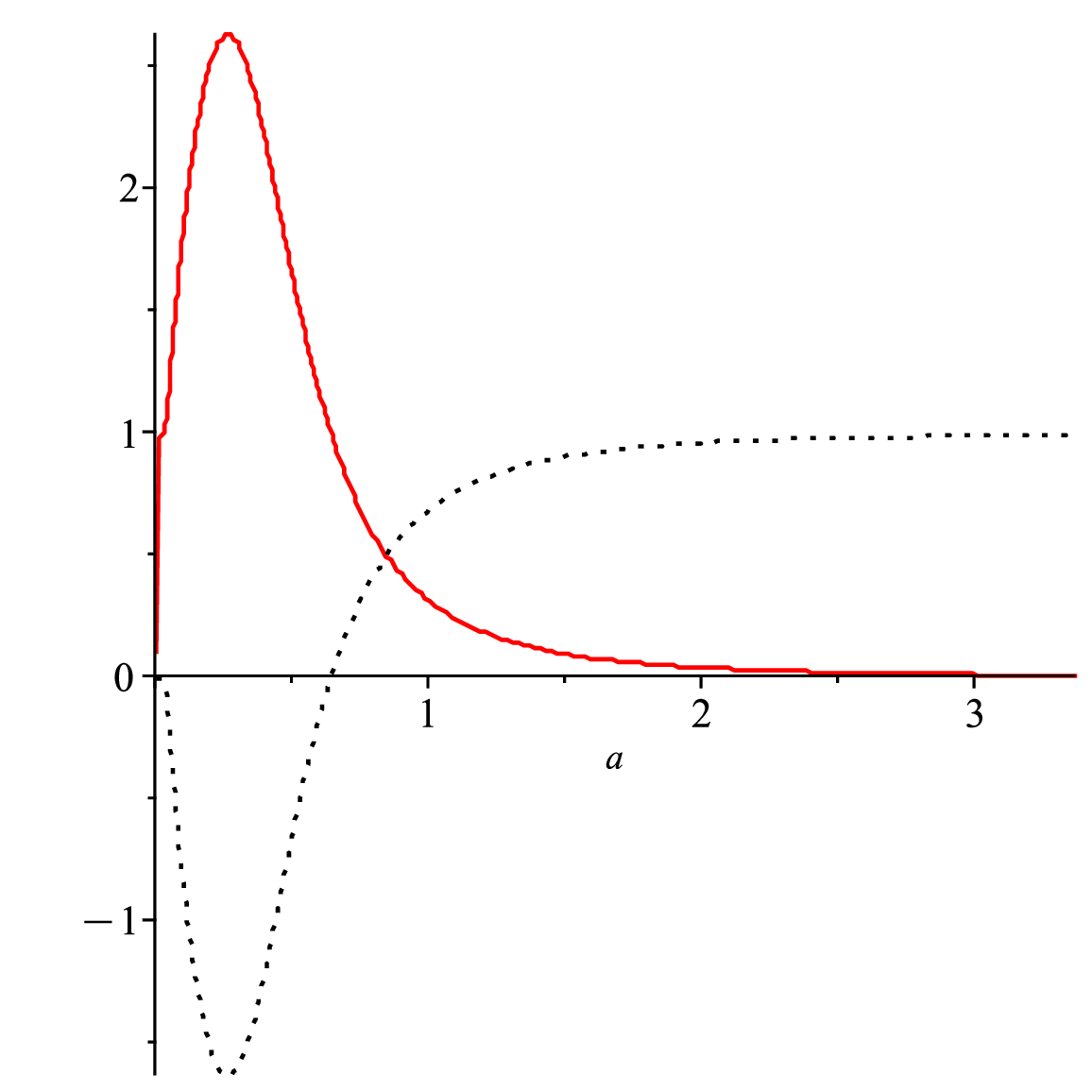}
\caption{The matter relative density, $\Omega_m$ (solid line), and dark energy relative density $\Omega_d$ (dotted line)
in terms of the scale factor.}
\label{fig3}
\end{figure}

Fig.(\ref{fig2}) shows that although the quintessence is activated in the radiation-dominated epoch,
the Universe enters a positive acceleration phase at $a\simeq 0.5$ and  $w(a=1)\simeq -0.7$.
The matter relative density $\Omega_m=\frac{\hat{\rho}_m}{3\hat{H}^2}=\frac{\rho_m}{3M_P^2H^2}$ is plotted in fig.(\ref{fig3}).
At $a=1$ we have $\Omega_m\simeq 0.32$ which is in the range reported in \cite{Planck} (based on $\Lambda CDM$) as $\Omega=0.315\pm 0.007$.
With the modified Friedmann equations, it is possible to have $\Omega_m>1$, as seen in the plot. In such a situation,
following (\ref{8}), we cannot specify an effective component with positive energy density as dark energy. $\Omega_d=\frac{\rho_d}{3M_P^2H^2}$
is plotted in fig(\ref{fig3}).
But at the present epoch $a=1$, we can use the equation (\ref{8}), and according to the fig.(\ref{fig3}) determine $\Omega_d(a=1)\simeq 0.68$.
 At $a=1$, this effective component's EoS parameter is $w_d=-1.03$. These results are compatible with \cite{Planck}, which,
 by combining Planck data with Pantheon supernova and Bao data, the EoS parameter of dark energy is obtained as $w_d=-1.03\pm 0.03$.

Now let us compute the GWS for the above model. In fig.(\ref{fig4}), we illustrate $\frac{c_T}{c}-1$ in terms of the scale factor.
\begin{figure}[H]
\centering
\includegraphics[height=6cm]{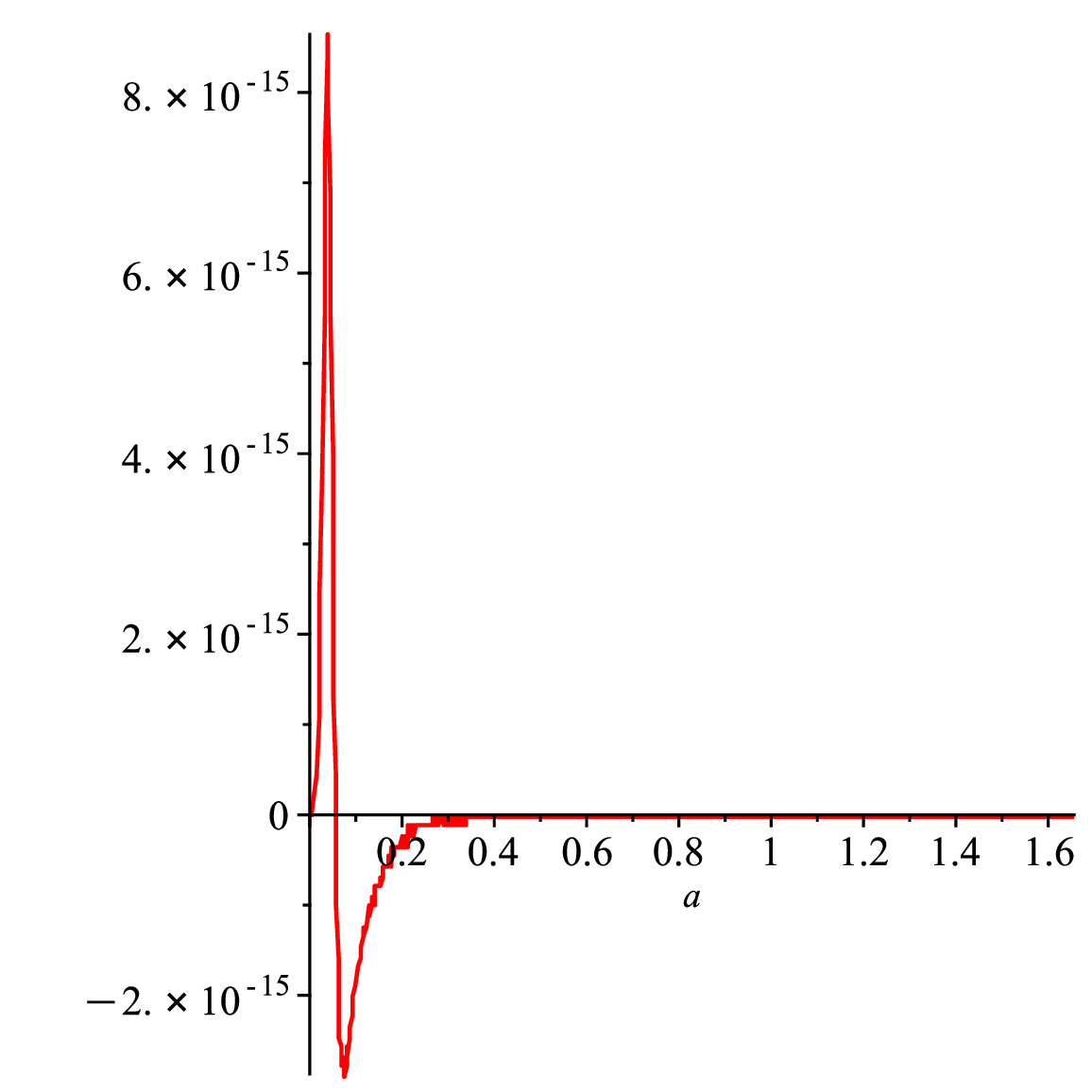}
\caption{ $\frac{c_T}{c}-1$ in terms of the scale factor.}
\label{fig4}
\end{figure}
For $a\simeq 1$ we have $\frac{c_T}{c}-1\simeq -7\times 10^{-17}$ in agreement with (\ref{15}).

Given the infinitesimal value of $\hat{\alpha}$, we expect the model to remain stable under tensor perturbations.
The avoidance of ghosts requires
\begin{equation}\label{25}
q_T=2G_4-2\dot{\phi}^2G_{4,X}+\dot{\phi}^2G_{5,\phi}-H\dot{\phi}^3G_{5,X}>0,
\end{equation}
which simplifies, after some computations, to
\begin{equation}\label{26}
q_T=-4H\dot{\phi}f_{,\phi}+u>0,
\end{equation}
and for this example this condition becomes:
\begin{equation}\label{27}
\frac{q_T}{M_P^2}=\hat{q}_T=4\hat{\alpha}^2\hat{H}\hat{\phi}\hat{\phi}'+\frac{1}{2}\epsilon^2\hat{\phi}^2+1>0.
\end{equation}
To avoid Laplacian instabilities, we also require $c_T^2>0$ which is satisfied.

For scalar perturbations, the relations are more complicated. Utilizing the results from \cite{stab5,stab6},
 and assuming $\frac{d^n f}{d\phi^n}=0, n\geq 3$, we find out the no-ghost and avoidance of Laplacian instability require
\begin{equation}\label{28}
q_s:=2(q_T+24H^4 f_{,\phi}^2)+3u_{,\phi}(u_{,\phi}-8H^2f_{,\phi})>0,
\end{equation}
and:
\begin{equation}\label{29}
c_s^2=-\frac{D^2c_T^2+2B D-2q_T}{q_s}>0,
\end{equation}
respectively, where
\begin{eqnarray}\label{30}
&&D=4H^2f_{,\phi}-u_{,\phi}\nonumber\\
&&B=4H^2(1+3w)f_{,\phi}+4u_{,\phi}
\end{eqnarray}
with $w=-1-\frac{2}{3}\frac{\dot{H}}{H^2}$ representing the effective EoS parameter of the Universe.
The quantities $\hat{q}_T=\frac{q_T}{M_P^2}$  and $\hat{q}_s=\frac{q_s}{M_P^2}$ and $c_s^2$ are plotted in fig.(\ref{fig4}).
\begin{figure}[H]
\centering
\includegraphics[height=6cm]{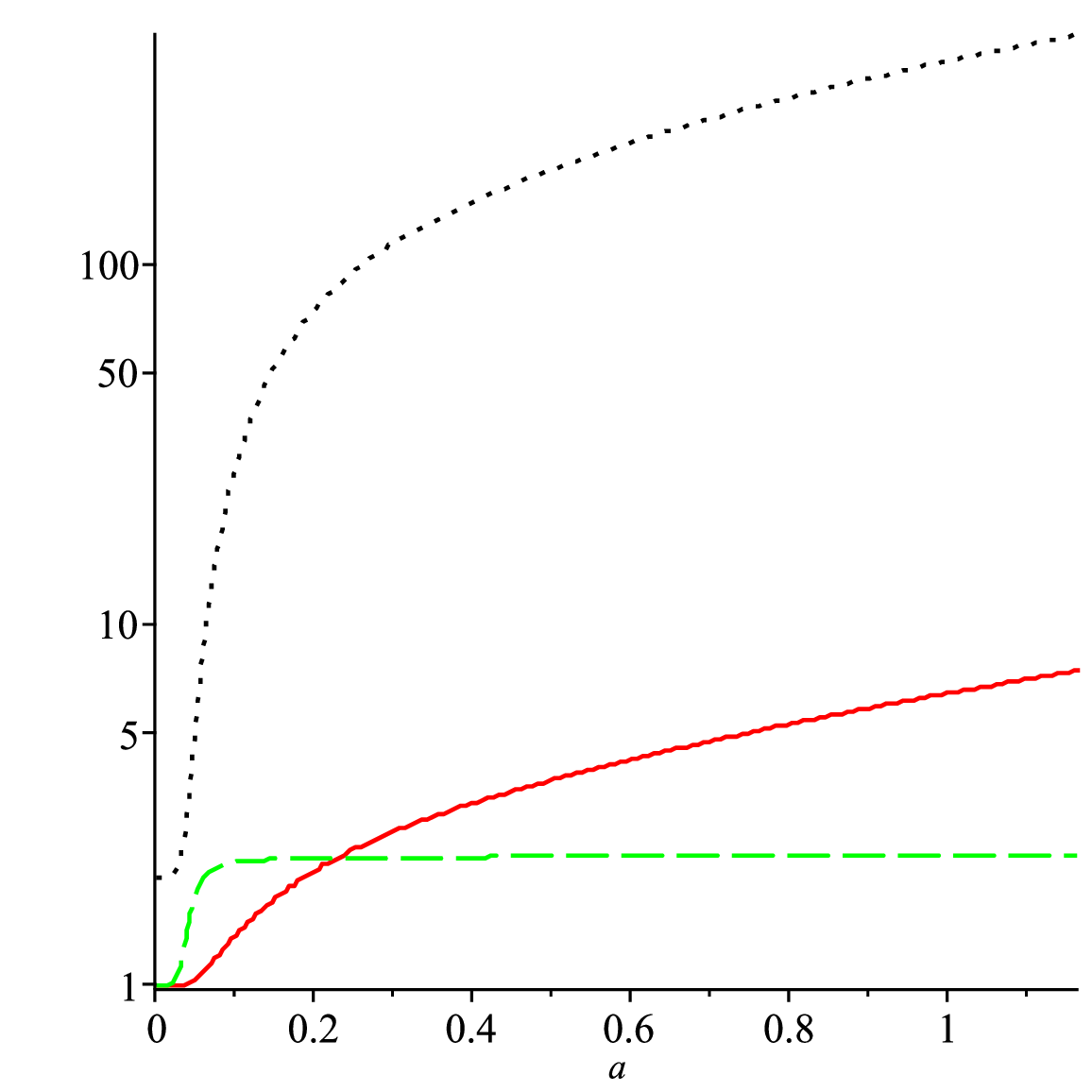}
\caption{$\hat{q}_T$ (solid line), $\hat{q}_s$ (dotted line), and  $c_s^2$ (dashed line) in terms of the scale factor.}
\label{fig4}
\end{figure}
This figure demonstrates that $q_T>0,\,\,q_s>0$ indicating the absence of ghosts. Moreover, since $c_s^2>0,\,\,\, c_T^2>0$,
there is no Laplacian instability.

The qualitative behavior depicted in the above figures arises directly from the intrinsic framework of our proposed model.
However, parameter adjustments have been made to ensure compatibility with the observational data reported in \cite{Planck}.
 Since our numerical analysis is formulated in terms of dimensionless parameters, any set ${\alpha, \rho_{\text{initial}}, \mu, H^*}$
 that results in the same ${\hat{\alpha}, \hat{\rho}_{\text{initial}}, \hat{\mu}}$ will produce identical results.

Before concluding, let us briefly examine the compatibility of our model, which exhibits a screening effect in dense regions
of FLRW space-time, with Solar system gravitational tests.

To study local gravitational test constraints on Horndeski's models, one may consider local regions such as the Solar System.
An effective way to assess the compatibility of these models with observations is through the parameterized post-Newtonian (PPN)
formalism \cite{Hoh}, by comparing theoretical predictions with observational constraints on the PPN parameters.
Specifically, the parameters are measured as
$\gamma=1+(2.1\pm 2.3)\times10^{-5}$  \cite{Ber}, and
$\beta=1+(-1.6\pm 1.8)\times 10^{-5}$ \cite{Gen}.
For the model described by (\ref{1}), assuming a static and spherically symmetric metric, the PPN parameters at first
post-Newtonian (1PN) order are given by \cite{Hoh}
\begin{equation}\label{r1}
\gamma(r)=\frac{2\omega +3-exp(-m_{\psi}r)}{2\omega +3+exp(-m_{\psi}r)},
\end{equation}
and
\begin{eqnarray}\label{r2}
&&\beta(r)=1+\frac{1}{(2\omega+3+e{-m_\psi r})^2}\bigg[\frac{\omega+\tau-4\omega\sigma}{2\omega+3}e^{-2m_\psi r} +(2\omega+3)m_\psi r\nonumber \\
&&\left(e^{-m_\psi r} \ln(m\psi r)-(m_\psi r +e^{m_\psi r})Ei(-2m_\psi r)-\frac{1}{2}e^{-2m_\psi}r\right)\nonumber \\
&&+\frac{6\tilde{\mu} r +3(3\omega+\tau+6\sigma+3)m_\psi^2r}
{2(2\omega+3)m_\psi}\Big(e^{m_\psi r}Ei(-3m_\psi r)\nonumber \\
&&-e^{-m_\psi r}Ei(-m_\psi r)\Big)\bigg],
\end{eqnarray}
where Ei denotes the exponential integral, and the effective mass $m_\psi$ is given by
\begin{equation}\label{r3}
m_\psi=\sqrt{\frac{-2G_{2(2,0)}}{G_{2(0.1)}-2G_{3(1,0)}+3\frac{G_{4(1,0)}^2}{G_{4(0,0)}}}},
\end{equation}
and
\begin{eqnarray}\label{r4}
&&\tau=\frac{G_{4(0,0)}^2}{2G_{4(1,0)}^3}
\left(G_{2(1,1)}-4G_{3(2,0)} \right) \nonumber \\
&&\sigma=\frac{G_{4(0,0)}G_{4(2,0)}}{G_{4(1,0)}^2}\nonumber\\
&&\omega=\frac{G_{4(0,0)}}{2G_{4(1,0)}^2}
\left(G_{2(0,1)}-2G_{3(1,0)}\right)\nonumber\\
&&\tilde{\mu}=\frac{G_{4(0,0)}^2G_{2(3,0)}}{G_{4(1,0)}^3}.
\end{eqnarray}
Here $G_{i(m,n)}$ denotes the partial derivatives of $G_i$ evaluated at the background values $(\phi_0,X_0)$
: $G_{i(m,n)}=\frac{1}{p!q!}\frac{\partial^{p+q}G_i}{\partial \phi^p \partial X^q}\mid_{\phi_0,X_0}$.
$\phi_0$, $X_0$ are given by $G_{2(0,0)}(\phi_0,X_0)=G_{2(1,0)}(\phi_0,X_0)=0$. In our model specified by (\ref{18}),
we have $G_3=0$ and,
\begin{eqnarray}
&&G_2=X-\frac{1}{2}\mu^2 \phi^2\nonumber\\
&&G_4=\frac{M_P^2}{2}+\frac{1}{4}\epsilon^2\phi^2+\alpha^2 X(2-\ln X).
\end{eqnarray}
Thus $\phi_0=X_0=0$.  From (\ref{r4}) we derive $\tau=\tilde{\mu}=0$ and  $\omega\to \infty$ and therefore
 $\gamma=1$ and $\beta=1$ \cite{Hoh}. This indicates that, within the standard PPN formalism and at the level of
 Solar System experiments, there is no deviation from General Relativity (GR).
However, due to the weak gravitational fields and low orbital velocities characterizing the Solar System, detecting non-linear
features and strong-field effects predicted by scalar-tensor theories requires more stringent tests.
Observations involving strong-field, highly dynamical systems—such as Extreme Mass-Ratio Inspirals (EMRIs),
 where a compact object spirals into a supermassive black hole—are better suited for probing potential
deviations from GR \cite{Ken}.

\section{Conclusion}

In this paper, we investigated the Scalar-Gauss-Bonnet model extended by incorporating an additional coupling of
the scalar field to the Ricci scalar curvature, situated within the broader framework of Horndeski theories.
Our study introduces a novel scalarization mechanism that results in late-time cosmic acceleration, starting
from an initial state with negligible dark energy density. Importantly, we ensured that the gravitational wave
speed remained consistent with the observational constraints provided by \cite{ob1, ob2, ob3}
(obtained from the nearly simultaneous detection of gravitational waves and gamma-rays from GW170817).

In our model, the Gauss-Bonnet (GB) term, while not directly contributing to the acceleration of the Universe at low redshifts,
plays a crucial role in stabilizing the quintessence field at a fixed point during the early universe. During this phase,
a $Z_2$ symmetry is maintained, ensures a vanishing dark energy density,
aligning with several previous studies \cite{Doran, Bartelmann, Hill, Gomez}.
As matter and radiation dilute, the Ricci scalar begins to dominate over the GB invariant,
initiating the evolution of quintessence through a cosmic scalarization process.
At early times {radiation era), the model mimics the behavior of the standard cosmological framework of General Relativity (GR).
However, after scalarization, the quintessence field becomes tachyonic, leading to non-trivial solutions that
diverge from GR and initiate late-time cosmic acceleration.

Our assumption regarding that the gravitational wave speed is within the bounds set by \cite{ob4},
suggests that scalarization occurs during the radiation-dominated epoch. However, due to the presence of a
friction term,
the emergence of dark energy from its initially dormant state is delayed until the matter-dominated epoch,
allowing it to eventually reach the same order of magnitude as the matter density in our present era,
despite differing dilution rates. Additionally, our stability analysis confirms that the model is free from
ghost and Laplacian instabilities. Using the results of \cite{Hoh}, We have also briefly examined the parameterized
post-Newtonian formalism in our model, which appears to satisfy solar system tests within this approximation.

In this study we have not studied the possibility of domain walls formation. This formation is due to spontaneous
symmetry breaking when the scalar field
acquires different non-zero vacuum expectation values in different regions of space.
However, In theories where $Z_2$ symmetry is broken spontaneously, domain wall formation can be prevented
through various mechanisms. If the scalar field begins with highly uniform initial conditions—naturally expected after
inflationary smoothing—it can transition coherently into a single vacuum state without creating topological defects \cite{Poly},\cite{Hint}.
Another approach involves adding a small explicit symmetry-breaking
term to the potential, which removes the degeneracy between vacua and selects a preferred ground state
universally \cite{Vil},\cite{Li}. Additionally, if the effective potential evolves adiabatically, allowing the scalar field to follow
the shifting minimum smoothly, spatial inhomogeneities are minimized, leading to uniform vacuum
selection \cite{Kh}, \cite{Br}. Also delay of the scalar field rolling until the matter dominated era (as in our model)
 may be important to avoid producing unwanted relics(like domain walls). This mechanism is very similar to moduli stabilization
 in axion cosmology \cite{ax}. These mechanisms collectively ensure the possibility to have a symmetry-breaking transition without generating
cosmologically problematic domain walls.

\textbf{Data Availability Statement}: The data generated or analyzed during
this study are included in the article.

\end{document}